\documentclass[aps,prl,twocolumn,showpacs,secnumarabic,nofootinbib,tightenlines,nobibnotes]{revtex4}
\usepackage{docs}
\usepackage{amsmath,amssymb,amsfonts,bm}
\usepackage{graphicx}
\usepackage{dcolumn}
\usepackage[colorlinks=true,linkcolor=blue]{hyperref}%

\begin{document}

\title{Surface Spectral Function of Momentum-dependent Pairing Potentials in a Topological Insulator: Application to Cu$_x$Bi$_2$Se$_3$}

\author{Liang Chen}
\author{Shaolong Wan}
\email[]{slwan@ustc.edu.cn}
\affiliation{Institute for Theoretical Physics and Department of Modern Physics\\
University of Science and Technology of China, Hefei, 230026, People's Republic of China}

\date{\today}

\begin{abstract}
We propose three possible momentum-dependent pairing potentials
for candidate of topological superconductor (for example
Cu$_x$Bi$_2$Se$_3$), and calculate the surface spectral function
and surface density of state with these pairing potentials.
We find that the first two can give the same spectral functions
as the fully-gapped and node-contacted pairing potentials given
in [Phys. Rev. Lett. 105, 097001], and that the third one can
obtain topological non-trivial case which exists flat Andreev
bound state and preserves the $C_3$ rotation symmetry. We hope our
proposals and results be judged by future experiment.
\end{abstract}

\pacs{74.20.Rp, 73.20.At, 74.20.Mn}

\maketitle

%-----------------------------------------------------------------
% The body of the paper
%-----------------------------------------------------------------

\indent Recently, topological insulators \cite{Hasanrmp2010,
QiRMP20, Moorenature2010} have attracted great attention in
condensed matter physics for their physical properties. The
experimental \cite{Konigscience2007, Hsiehnature2008,
Hsiehnature2009, YXianature2009, Hsiehprl2009, YLChenscience2009}
and theoretical \cite{Fuprl2007, Fuprb2007, Mooreprb2007,
Qiprb2008, HJZhangnature2009} investigations show that topological
insulators (TI) are fully gapped in the bulk but gapless on the
surface, which is protected by time-reversal symmetry. And these
surface states indicate the massless Dirac fermions. Then, the
researches on topological insulators are generalized to on
topological superconductors (TSC) \cite{SchnyderPRB2008,
kitaevAIP2009, WenPRB2012} soon. Similarly, researches show that
topological superconductors also have gapless surface states,
which indicate massless Majorana fermions \cite{QiPRL.102.187001},
and that the property on TSC is protected by topological bulk
properties and characterized by a topological invariant
\cite{SchnyderPRB2008, QiPRB.81.134508}. As we know, the Majorana
fermions are of great interest in fundamental physics and have
potential applications in quantum computation
\cite{FuPRL.100.096407, WilczekNatureP2009}.

\indent The experiment \cite{HorPRL.104.057001} finds that a
superconductive phase is induced at transition temperature $T_c =
3.8 K$ when copper atoms are doped into topological insulator
Bi$_2$Se$_3$ with the concentration of Cu in range 0.12$\sim$0.15.
And the work \cite{WrayNatureP2010855} shows that the surface
state of Cu$_x$Bi$_2$Se$_3$ is topological non-trivial. A recent
experiment \cite{SasakiPRL2011} further confirms the existence of
topological surface state by measuring the surface density of
states (SDOS). However, a more detailed analysis show that a
gapless and topological non-trivial bulk band structure may be
preferred.

\indent Up to now, the exact pairing mechanism still be unclear,
but there are some theoretical proposals on the pairing symmetry
of Cu$_x$Bi$_2$Se$_3$, including odd-parity pairing potential
proposed by Fu and Berg \cite{FuPRL.105.097001} and Sato
\cite{PRB.79.214526,PRB.81.220504} and the explanation of Sasaki
{\it et.al.}\cite{SasakiPRL2011} according to experimental
results. However, in all theoretical proposals, they assume that
the pairing potentials are momentum-independent. For this reason,
we ask whether we can search for some pairing symmetries which
induce new topological surface state and act as candidates of
pairing symmetry of Cu$_x$Bi$_2$Se$_3$, if the pairing potentials
are momentum-dependent. Further more, we know that the $\Delta_4$
suggested by Ref. \cite{SasakiPRL2011} breaks $C_3$ rotation
symmetry of the rhombohedral lattice. So we also want to know
whether there is possible pairing potential which is topological
non-trivial, node-contacted and preserves the $C_3$ rotation
symmetry. In order to search for answers of these questions, in
this paper, we propose three momentum-dependent pairing potentials
for Cu$_x$Bi$_2$Se$_3$, and calculate the surface spectral
functions with these pairing potentials. And we find that the
first two sorts of our pairing potentials are similar to
$\Delta_2$ and $\Delta_4$ proposed in Ref. \cite{FuPRL.105.097001}
and can obtain what $\Delta_2$ and $\Delta_4$ give, and that the
third one can get topological non-trivial case which exists flat
Andreev bound state (ABS) and preserve the $C_3$ rotation
symmetry.

\indent As reported \cite{WrayNatureP2010855}, near the
$\Gamma$-point, the band dispersion of normal state of
Cu$_x$Bi$_2$Se$_3$ can be described by the low energy effective
Hamiltonian \cite{HJZhangnature2009} for Bi$_2$Se$_3$, with a
finite chemical potential in the conduction band induced by copper
doping. The Hamiltonian is
\begin{equation}
\label{eq1}
h(\bm{k})=(M\tau_{z}-\mu)+\tau_{x}\left(A(k_x\sigma_{x}+k_y\sigma_y)+{B}k_z\sigma_z\right),
\end{equation}
where $M$ is the rest mass, $\mu$ is the chemical potential, $A$
and $B$ are Fermi velocity along different directions,
$\tau_{z}=\pm1$ denotes the two orbits, and $\sigma_{x,y,z}$ are
spin Pauli matrices. For the superconductivity phase, the
Hamiltonian can be written in the Bogoliubov-de Gennes (BdG)
formalism:
\begin{equation}
\label{eq2}
\mathcal{H}=\left(
              \begin{array}{cc}
                h(\bm{k}) & \Delta(\bm{k}) \\
                \Delta^{\dagger}(\bm{k}) & -h^{T}(-\bm{k}) \\
              \end{array}
            \right),
\end{equation}
where $\Delta(\bm{k})$ is the pairing potential ($4\times 4$
matrix). For the time reversal invariant case, the pairing
potential can be divided into two parts according to inversion
symmetry:
\begin{eqnarray}
\Delta(\bm{k}) &=& \Delta_{1}(\bm{k}) + \Delta_{2}(\bm{k}),\label{eq3}\\
\Delta_{1}(\bm{k})&=&-i k_{j}\sigma_y\left(\sigma_{\alpha}\Delta_{1,s}^{\alpha{j}}+\tau_{z}\sigma_{\alpha}\Delta_{1,as}^{\alpha{j}}\right),\label{eq4}\\
\Delta_{2}(\bm{k})&=&k_{j} \left(\tau_{x}\sigma_{\alpha}\Delta_{2}^{\alpha{j}}+i\tau_{x}\Delta_{2}^{0j} \right),\label{eq5}
\end{eqnarray}
where $\Delta_{1,s}^{\alpha{j}}$, $\Delta_{1,as}^{\alpha{j}}$,
$\Delta_{2}^{0j}$ and $\Delta_{2}^{\alpha{j}}$ are real functions
of momentum and inversion symmetric, the summation convention is
used in this paper. One can find that
$\Delta_{2}(\bm{k})=\hat{\mathcal{P}}^{-1}\Delta_{2}(\bm{k})\hat{\mathcal{P}}$
is inversion symmetric and
$\Delta_{1}(\bm{k})=-\hat{\mathcal{P}}^{-1}\Delta_{1}(\bm{k})\hat{\mathcal{P}}$
is inversion anti-symmetric, where the inversion operator is
$\hat{\mathcal{P}}=\tau_{z}$ in coordinate space. As we know, if
the pairing potential is dominated by $\Delta_{1}(\bm{k})$ and
fully gapped in the Brillouin zone, the criterion for topological
odd-parity superconductor
\cite{FuPRL.105.097001,PRB.81.220504,PRB.79.214526} claims that
the system is topological nontrivial.

In the following, we consider three typical cases of
$\Delta_1(\bm{k})$ and calculate the surface spectral function
with them. In order to calculate the spectral function
numerically, we consider a lattice model which the low energy
effective Hamiltonian is Eq.(\ref{eq1}). For the normal state of
the Hamiltonian, we use the model and parameters given in
supplemental material of Ref. \cite{SasakiPRL2011}.

In the first case, for superconductive potential,
Eqs.(\ref{eq3})-(\ref{eq5}), we consider
$\Delta_{1,as}^{\alpha{j}} = \Delta_2^{\alpha{j}} =
\Delta_2^{0j}=0$ and $\Delta_{1,s}=\Delta\text{diag}(A, -A, B)$,
as an example, and other cases are similar. In this case, the
pairing potential takes
\begin{equation}
\label{eq6}
\Delta(\bm{k})=\Delta\left(
\begin{array}{cc}
 -A\left(k_x-{i}k_y\right) & B k_z\\
 {B}k_z & A\left(k_x+{i}k_y\right)
\end{array}\right)\otimes\tau_{0},
\end{equation}
where $\Delta$ is a dimensionless parameter determined by energy
gap of superconductivity, $\tau_0$ is a $2\times2$ identity matrix
in orbital space. One of compactifications of Eq.(\ref{eq6}) can
be given as
\begin{equation}
\label{eq7}
\Delta(\bm{k})=\Delta\left(
\begin{array}{cc}
 -A_2^{-} & A_1 \\
 A_1 & A_2^{+}
\end{array}\right)\otimes\tau_{0},
\end{equation}
where we refer to the definition of $A_1$, $A_2^{\pm}$ in the
supplemental Material of Ref. \cite{SasakiPRL2011}, and take the
same parameter values of Ref. \cite{SasakiPRL2011} in our
numerical calculation. In this case, the lattice model can
preserve the same translation symmetry of the discrete version of
$h(\bm{k})$ and turn to Eq.(\ref{eq6}) in the low energy limit
($\bm{k}\rightarrow0$). We must point out that this lattice model
to reproduce the low energy effective Hamiltonian is not valid for
$k\gg{k}_F$.

\indent By using the method in Refs. \cite{WangPRB.81.035104,
HaoPRB.83.134516}, we can obtain the surface spectral function.
Considering a semi-infinite system which has a surface at $z=0$,
the momentum which parallels to the surface
$\bm{k}_{||}=(k_x,k_y)$ is a good quantum number, and the
partition function of the system with an open surface at $z=0$ can
be written as:
\begin{eqnarray}
\label{eq8}
\mathcal{Z}&=&\int\mathcal{D}\psi^{\dagger}\mathcal{D}\psi \text{exp}\left\{
i\int{dt}\sum_{\bm{k}_{||}}\sum_{n=0}^{\infty}\left[
\psi_n^{\dagger}(i\partial_{t}-H_{\bm{k}_{||}})\psi_n
\right. \right.\nonumber \\
& &\left. \left.
 + \left(\psi_{n}^{\dagger}V_{\bm{k}_{||}}\psi_{n+1}+h.c.\right)
\right]
\right\},
\end{eqnarray}
where $\psi_n$ is the wave function for the $n$th layer,
$H_{\bm{k}_{||}}$ is the intralayer Hamiltonian, $V_{\bm{k}_{||}}$
is the interlayer coupling, and $h.c.$ means Hermitian conjugate.
The recursive integration layer by layer gives the following
Green's function for the surface state:
\begin{equation}
\label{eq9}
G^{-1}(\bm{k}_{||},\omega) = G_0^{-1}(\bm{k}_{||},\omega) - V^{\dagger}_{\bm{k}_{||}}G^{-1}(\bm{k}_{||},\omega)V_{\bm{k}_{||}},
\end{equation}
where
$G_0(\bm{k}_{||},\omega)=\left(\omega-H_{\bm{k}_{||}}\right)^{-1}$
is the free Green's function without interlayer coupling. The
Green's function of surface state is calculated by the quick
iterative scheme \cite{Sancho0305-4608-14-5-016} for $T$-matrix.
Finally, the surface spectral function is given in the following
form,
\begin{equation}
\label{eq10}
A(\bm{k}_{||},\omega) = -\frac{1}{\pi} \text{Im}\,\text{Tr} G(\bm{k}_{||},\omega).
\end{equation}
One can also calculate the SDOS by integrating $A(\bm{k}_{||},\omega)$ over momentum,
\begin{equation}
\label{eq11}
\rho_{s}(\omega) = \int \frac{d^2\bm{k}_{||}}{(2\pi)^2} A(\bm{k}_{||},\omega).
\end{equation}
In order to agree with the moment independent pairing, in the
calculation, we make the dimensionless parameter $\Delta$ be
$0.15$, the maximum gap size be about $0.05\text{eV}$, and the
truncation of momentum $|k_x|=|k_y|$ be $0.6\text{eV}$, because,
as pointed above, the lattice model is not valid for $k\gg{k}_F$.

\begin{figure}
  % Requires \usepackage{graphicx}
  \includegraphics[width=8cm]{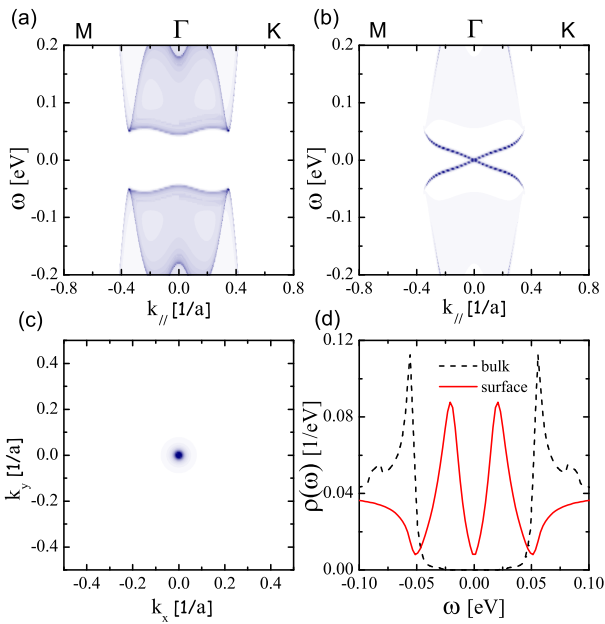}
  \caption{Model calculation of bulk (a) and surface (b) spectral function $A(\bm{k},\omega)$
  for BdG Hamiltonian with pairing potential $\Delta_{1,s}=\Delta\text{diag}(A, -A, B)$.
  (c) Surface spectral function as a function of momentum for $\omega=0$.
  (d) Bulk (black dash line) and surface (red solid line) density of state(DOS). The false color mappings of $A(\bm{k},\omega)$ in
  (a), (b) and (c) are in arbitrary units. Parameters for model calculation have been given in the
  context.}\label{fig1}
\end{figure}

\indent Now we consider the surface spectral function for some
special pairing symmetries. In the first case, the pairing
potential is given in Eq.(\ref{eq6}), which looks like a direct
sum of two pairing potentials to describe the ${}^3$He-B phase.
Because, in topological insulator, strong spin-orbit coupling
between different orbits makes the pairing potentials between
different energy bands be complicated, we must calculate the
topological invariant carefully. We can identify the topological
invariant by the criterion for topological odd-parity
superconductor \cite{FuPRL.105.097001,PRB.81.220504,PRB.79.214526}
which shows that the pairing symmetry of Eq.(\ref{eq6}) is
topological non-trivial and we can also calculate the winding
number directly \cite{SchnyderPRB2008},
\begin{equation}
\label{eq12} N_w = \frac{1}{24\pi^2}\int d^3\bm{k} \epsilon^{ijk}
\text{Tr}[Q_{\bm{k}}^{\dagger}\partial_{i}Q_{\bm{k}}Q_{\bm{k}}^{\dagger}\partial_{j}Q_{\bm{k}}Q_{\bm{k}}^{\dagger}\partial_{k}Q_{\bm{k}}],
\end{equation}
where $Q_{\bm{k}}=2P_{\bm{k}}-1$,
$P_{\bm{k}}=\sum_{n\in{occ}}|u_{n}(\bm{k})\rangle\langle{u}_{n}(\bm{k})|$
is the projector onto the occupied Bloch states. We use the later
method and find that the winding number is totally determined by
the topology of {\it Fermi surface} and
$N_w=-1\text{~Sgn}(\Delta)$ if the pairing potential (\ref{eq6})
is non-vanishing only for a thin spherical shell around the {\it
Fermi momentum} $k\sim{k}_F$, which implies that although the
pairing potential is written in four bands (different spins and
different orbitals), it can be continuously deformed to the {\it
weak pairing limit} \cite{QiPRB.81.134508} on the Fermi surface.
The spectral function for the Hamiltonian with pairing potential
(\ref{eq6}) (as shown in Fig.\ref{fig1}) shows that the bulk state
is fully gapped (as shown in Fig.\ref{fig1}(a)) and there is an
Andreev bound state on the surface (as shown in
Fig.\ref{fig1}(b)). Similar to the momentum-independent odd-parity
pairing potential $\Delta_{2}$ in Ref. \cite{FuPRL.105.097001},
this pairing potential is inconsistent with experimental results
and there is a minimal of the SDOS at $\omega=0$
(Fig.\ref{fig1}(d)), which is not observed at the zero-bias
conductance peak. One of the explanation indicates that the bulk
band structure is topological non-trivial but with some point
nodes, which will induce a flat dispersion of surface helical
Majorana fermions and contribute a non-vanishing peak of SDOS at
zero energy. Among all sorts of momentum-dependent pairing
symmetries, there exist some species which possess this property,
resembling the $\Delta_{4}$ in Ref. \cite{FuPRL.105.097001}.

\begin{figure}
  % Requires \usepackage{graphicx}
  \includegraphics[width=8cm]{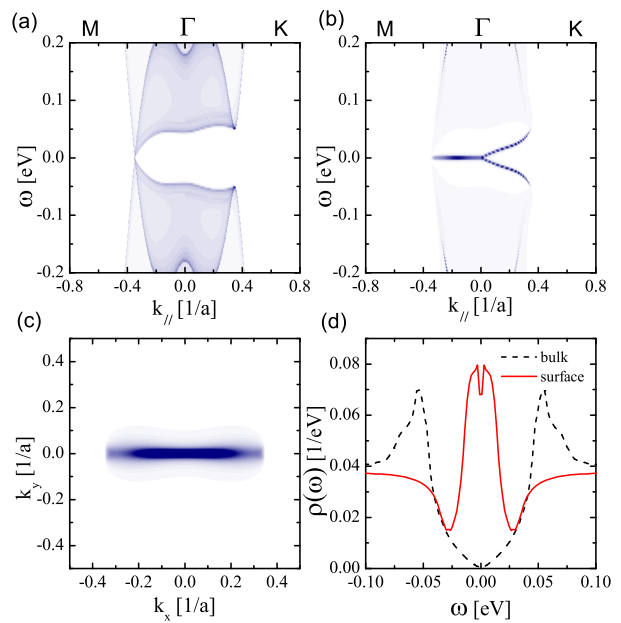}
  \caption{Model calculation of bulk (a) and surface (b) spectral function $A(\bm{k},\omega)$
  for BdG Hamiltonian with pairing potential $\Delta_{1,s}=\Delta\text{diag}(A, 0, B)$.
  (c) Surface spectral function as a function of momentum for $\omega=0$.
  (d) Bulk (black dash line) and surface (red solid line) DOS. The false color mappings of $A(\bm{k},\omega)$
  in (a), (b) and (c) are in arbitrary units. Parameters for model calculation have been given in the context.} \label{fig2}
\end{figure}

In the second case, we consider the pairing potential given by
Eq.(\ref{eq4}) with $\Delta_{1,s}=\Delta\text{\,diag}(A,0,B)$ and
$\Delta_{1,as}^{\alpha{j}}=0$ as other example, and find that the
bulk bands have two point nodes at
$\bm{k}=(0,\pm\sqrt{\mu^2-M^2}/A, 0)$, and the $Q$-matrix for this
pairing potential is well defined in the Brillouin zone excluded
these two points. In the weak pairing limit, the $Q$-matrix can be
written as
\begin{equation}
\label{eq13}
Q_{\bm{k}} = -\frac{i}{2}\,\text{Sgn}(\Delta)[\cos(\theta)\sigma_{z}+\sin(\theta)\sigma_{x}]\otimes(\tau_{0}-\tau_{z}),
\end{equation}
near the two singularity points on the Fermi surface, where
$\theta=\arctan[{A}k_x/({B}k_z)]$. Here, we have made a unitary
transformation to express $Q_{\bm{k}}$ in the eigenvalue
representation of $h(\bm{k})$ and the $\sigma_{x,z}$ and
$\tau_{0,z}$ in the same form as before but with different
meanings. Eq.(\ref{eq13}) shows that there are two ABS on the
boundary of $xz$-plane, which is similar to the chiral $p$-wave
superconductor but time reversal symmetry is unbroken here, and
these ABS are stable if the two point nodes are disconnected. The
spectral function of this pairing potential is given in
Fig.\ref{fig2}. The SDOS has a non-vanishing value at $\omega=0$
(Fig.\ref{fig2}(d)).

\indent However, this pairing symmetry seems to be also
unlikelihood, because there are only two point nodes of the bulk
bands which breaks the symmetry of $D_{3h}$ group. As shown in
Fig.\ref{fig2}(c), the spectral function of surface state for
$\omega=0$ is not invariant under the $C_{3}$ rotation operation
in $k_{x}k_{y}$-plane. As we know, the $\Delta_4$ suggested for
the pairing symmetry of Cu$_x$Bi$_2$Se$_3$ in Ref.
\cite{SasakiPRL2011} also exists such a deficiency.

In order to solve this question, in the third case, we ask to
construct a pairing potential which must satisfy the following
conditions, (1) it is topological non-trivial, (2) its band
structure has some point nodes in the $k_{x}k_{y}$-plane and (3)
its spectral function preserves the $C_{3}$ rotation symmetry, and
find that the high order terms of momentum are indispensable.

\begin{figure}
  % Requires \usepackage{graphicx}
  \includegraphics[width=8cm]{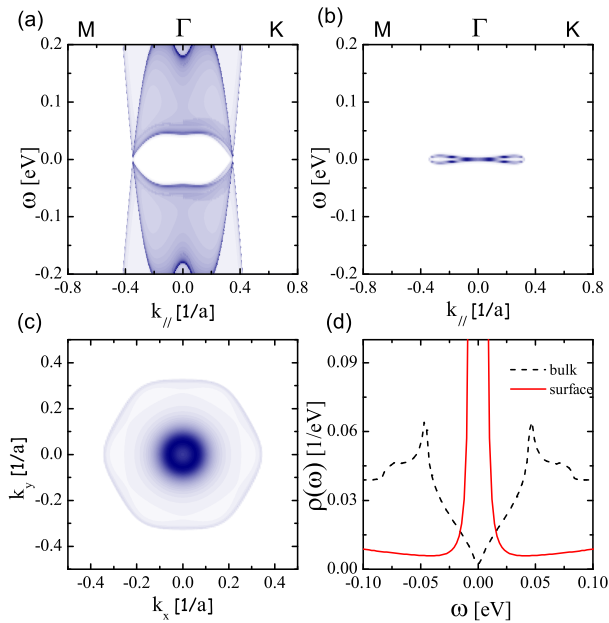}
  \caption{Model calculation of bulk (a) and surface (b) spectral function $A(\bm{k},\omega)$
  for BdG Hamiltonian with pairing potential given in Eq.(\ref{eq14}).
  (c) Surface spectral function as a function of momentum for $\omega=0$.
  (d) Bulk (black dash line) and surface (red solid line) DOS. The false color mappings of $A(\bm{k},\omega)$ in (a), (b) and (c) are in arbitrary units.
  Parameters for model calculation have been given in
  the context.} \label{fig3}
\end{figure}

\indent Enlightened by the hexagonal warping effects
\cite{FuPRL.103.266801} of the surface state of topological
insulator, we construct the following pairing potential,
\begin{equation}
\label{eq14}
\Delta_{1}(\bm{k}) = \Delta [B{k}_z + \lambda A^3(k_{x}^{3}- 3k_{x} k_{y}^{2}) ] \sigma_{x}\otimes\tau_{0},
\end{equation}
where $\lambda$ is a parameter and $\lambda=2eV^{-2}$ in
calculation, for simplicity, we also choose
$\Delta_{1,as}^{\alpha_j}=\Delta_{2}(\bm{k})=0$, as another
example. The term proportional to $B k_{z}$ is applied to open a
gap at the $\Gamma$-point, which can be replaced by the $\Delta_3$
in Ref. \cite{FuPRL.105.097001} or some other topological
non-trivial terms. All of them can give the similar spectral
functions. In addition, we must emphasize that the $k^3$ terms are
not high order corrections and they are as important as the linear
order terms of momentum for the weak pairing limit. The spectral
function and SDOS are given in Fig.(\ref{fig3}), now $C_{3}$
rotation symmetry is preserved (as shown in Fig.\ref{fig3}(c)),
the net effect of flat helical Majorana fermions induced by six
point nodes of bulk bands accumulates a sharp surface spectral
function peak around the $\Gamma$-point for $\omega=0$, and this
effect is also manifested in the SDOS (as shown in
Fig.\ref{fig3}(d)).

\indent Finally, we discuss the other choice in
Eqs.(\ref{eq3})-(\ref{eq5}). In general case, the pairing symmetry
$\Delta_{1,as}^{\alpha{j}}$ of the anti-symmetric orbits is
similar to the pairing symmetry $\Delta_{1,s}^{\alpha{j}}$ of the
symmetric orbits, we can construct parallel theories for pairing
potentials which are similar to above examples or their
combination. For the fully bulk-gapped systems, we find that they
can deform to each other continuously. The difference between
$\Delta_{1,as}^{\alpha{j}}$ and $\Delta_{1,s}^{\alpha{j}}$ can not
be distinguished by the shape of spectral function. In addition,
after calculating the spectral functions for $\Delta_{2}(\bm{k})$
at the linear order of momentum, we find that the system is bulk
gapless when $\Delta_{2}^{0j} \ne 0$ and others are zero, and the
Hamiltonian with only $\Delta_{2}^{\alpha{j}} \ne 0$ can be bulk
gapped and topological trivial, its winding number $N_w=0$.

\indent In summary, we calculate the spectral function and SDOS
for three typical momentum-dependent pairing potential in a
topological insulator, which may act as the candidate for the
pairing symmetry of superconductor Cu$_x$Bi$_x$Se$_3$, we find
that similar to momentum-independent $\Delta_2$ and $\Delta_4$
proposed in Ref. \cite{FuPRL.105.097001}, the pairing potentials
of the momentum-dependent also permit ABS induced by topological
non-trivial fully-gapped or node-contacted bulk bands, as shown in
the first and second cases. We point out that the previous
topological non-trivial node-contacted pairing potentials do not
preserve the $C_{3}$ rotation symmetry of lattice structure, and
we find a solution for this inconsistence, in the third example.

\indent We must clarify that the node-contacted bulk band
structure is not the unique explanation for zero-bias conductance
peak, a recent paper \cite{HsiehPRL.108.107005} shows that a fully
gapped bulk state with a twisted dispersion of ABS is also
possible. We hope these different pairing symmetries be judged by
future experiment and be helpful for the study of the pairing
mechanism of Cu$_x$Bi$_2$Se$_3$.

\indent {\it Acknowledgement:} This work is supported by NSFC
Grant No.10675108.

%-----------------------------------------------------------------
% Sec**: References
%-----------------------------------------------------------------
%\nocite{*}
%\bibliographystyle{apsrev4-1}
\bibliography{reference}

\end{document}